# Quantum simulation approach to ultra-weak magnetic anisotropy in a frustrated spin-1/2 antiferromagnet

Ki Won Jeong[*], Jae Yeon Seo[*], Sunghyun Lim, Jae Min Hong, Hyeon Jun Ryu, Jongseok Byeon, Kyungsun Moon, Nara Lee, and Young Jai Choi

Department of Physics, Yonsei University, Seoul 03722, Korea

**Abstract**

The intrinsic equivalence between electron spin and qubit offers a natural foundation for quantum simulations of magnetic materials. However, incorporating magnetocrystalline anisotropy (MCA), a key feature of real magnets, remains a major challenge. Here, we develop a quantum simulation framework for MCA in $CuSb_2O_6$, a spin-1/2 antiferromagnet with alternating ferromagnetic chains arising from frustrated, anisotropic exchange interactions in a nearly square lattice. The $Cu^{2+}$ spin network is modeled as a four-qubit square lattice, with four paired ancilla qubits introduced to encode angle-dependent MCA. This two-qubit representation per spin site resolves the limitation that squared Pauli operators yield only the identity, enabling MCA terms to be faithfully embedded into quantum circuits. Using the variational quantum eigensolver, we determine an exceptionally small easy-axis MCA constant, just 0.00022% of the nearest-neighbor exchange interaction, yet sufficient to drive a spin-flop transition with 90° spin reorientation and strong angular variation in magnetic torque. Beyond this regime, the simulations uncover a half-saturated magnetic phase at ultra-high fields, stabilized by anisotropic next-nearest-neighbor interactions. Our findings demonstrate the feasibility of resource-efficient quantum simulations of complex magnetic phenomena in real materials.

[*]These authors contributed equally to this work.
Correspondence: Kyungsun Moon (kmoon@yonsei.ac.kr), Nara Lee (eland@yonsei.ac.kr) or Young Jai Choi (phylove@yonsei.ac.kr)

**Introduction**

Quantum computing is rapidly emerging as a next-generation platform for simulating quantum systems, offering intrinsic advantages in representing quantum states and their evolution.[1-6] Among hybrid quantum-classical algorithms, the variational quantum eigensolver (VQE) has shown particular promise by variationally optimizing quantum circuits to approximate ground-state energies.[7-10] While VQE is compatible with today's noisy intermediate-scale quantum (NISQ) hardware, it can also be implemented on idealized simulators, permitting high-fidelity studies of problems that remain challenging for current devices.[8,11]

While VQE has demonstrated success in quantum chemistry,[9,12-15] spin systems provide an even more natural application, as qubits inherently represent spin-1/2 degrees of freedom without requiring fermion-to-qubit mappings.[16-18] This direct correspondence simplifies circuit construction and interpretation. Many spin-1/2 magnets exhibit ordered spin arrangements such as chains,[19-21] ladders,[22-24] or two-dimensional lattices,[25-27] allowing efficient simulation with modest qubit counts.

Despite this potential, extending quantum simulations to real magnetic materials has remained limited, primarily due to the difficulty of incorporating magnetocrystalline anisotropy (MCA), a key property that governs spin orientation relative to the crystal lattice and its response to external magnetic fields.[28,29] MCA introduces angle-dependent interactions that are typically nonlinear and challenging to encode using standard gate-based quantum circuits. Nevertheless, accurate modeling of MCA is essential, as it plays a critical role in spin reorientation phenomena, such as spin-flop transitions, commonly observed in uniaxially anisotropic antiferromagnets.[30-32]

In this context, spin-1/2 antiferromagnets, especially those based on $Cu^{2+}$ magnetic ions, offer ideal testbeds for quantum simulation. These materials exhibit well-defined anisotropic spin structures and sharp magnetic phase transitions that can be directly compared with quantum simulation outputs, providing a unique opportunity to validate such approaches with experimental results.[33-35]

In this work, we implement a structure-adapted quantum circuit for $CuSb_2O_6$, a square-lattice spin-1/2 antiferromagnet with frustrated spin chains aligned along the *b* axis.[36-38] Using four system qubits to represent physical spins and four ancilla qubits to encode angle-dependent MCA terms, we execute the circuit on the Qiskit statevector simulator within the VQE algorithm. The simulations reproduce the 90° spin-flop reorientation and quantify the extremely small anisotropy constant (0.00022% of the dominant exchange interaction). These results are validated against magnetization and, more sensitively, angular torque measurements, which directly probe the collapse and recovery of the spin-flop state. Extending to ultra-high fields, the simulations further reveal a previously hidden half-saturated magnetic phase sustained by anisotropic next-nearest-neighbor couplings. These findings demonstrate how quantum simulation can illuminate subtle interactions and field-induced phases in real materials, establishing a generalizable route for embedding MCA into quantum simulations of magnetism.

**Results**

**Frustration-driven spin chains and quantum simulation in a nearly square lattice**

$CuSb_2O_6$ crystallizes in a trirutile-type structure, where $Cu^{2+}$ ions form a square lattice.[36] At 380 K, the tetragonal phase transforms to a monoclinic $P2_1/n$ structure (Fig. 1a) with lattice parameters $a$ = 0.46401 nm, $b$ = 0.46411 nm, $c$ = 0.93077 nm, and $\beta$ = 91.113°.[36,39] This transition generates crystallographic twins,[36,38] stemming from two energetically equivalent orientations of the Jahn-Teller-distorted $CuO_6$ octahedra (Supplementary Fig. S1). This structural context provides the foundation for the distinct exchange pathways and frustration effects described below.

In the low-temperature ($T$) phase, $Cu^{2+}$ ions form a nearly square sublattice that develops long-range magnetic order at $T_N$ = 8.5 K,[40] as evidenced by anisotropic susceptibility (Supplementary Fig. S2). Nearest-neighbor (NN) spins are ferromagnetically coupled along the lattice edges, with slightly different strengths along the $b$ and $a$ axes ($J_1 < 0$ and $J_1' < 0$). Next-nearest-neighbor (NNN) antiferromagnetic interactions act along the diagonals: a direct $Cu^{2+}$–$O^{2-}$–$O^{2-}$–$Cu^{2+}$ bridge produces stronger coupling ($J_2 > 0$), while its absence on the opposite diagonal makes $J_2'$ significantly weaker.[41,42] This interplay of ferromagnetic NN and antiferromagnetic NNN couplings leads to spin frustration within the square lattice.

In an ideal square lattice where $J_1 = J_1'$, this generates two degenerate ground states with ferromagnetic chains aligned along either the $a$ or $b$ axis (Supplementary Fig. S3). In real $CuSb_2O_6$, however, subtle structural anisotropy lifts this degeneracy.[38,43] Although the $b$-axis lattice constant is only slightly larger than $a$,[36,40] the $Cu^{2+}$–$O^{2-}$–$Cu^{2+}$ bonding angle along $b$ (109.6°) is closer to 90° than along $a$ (112.8°) (Fig. 1b). By the Goodenough-Kanamori rules,[44,45] this strengthens ferromagnetic exchange along $b$ ($|J_1| > |J_1'|$), favoring chain alignment along $b$, consistent with neutron diffraction results.[41] Unlike the geometric frustration typical of triangular or kagome lattices, this rare case in a nearly square lattice arises from competing, direction-dependent exchanges, making it a distinctive feature of $CuSb_2O_6$.

To capture frustration and anisotropic exchange in $CuSb_2O_6$, we construct a quantum simulation framework that maps each $Cu^{2+}$ ion, effectively a spin-1/2 due to its $3d^9$ electronic configuration stabilized by a Jahn–Teller distortion, onto a qubit (Fig. 1c). Within a VQE scheme, the magnetic Hamiltonian is encoded into Pauli operators, and a parameterized Ansatz represents spin orientations on the Bloch sphere, providing a compact and scalable approach for simulating complex spin textures. A key remaining challenge is how to encode MCA, whose angular dependence cannot be retained within conventional single-qubit mappings.

**Encoding magnetocrystalline anisotropy for quantum simulation**

As shown in Fig. 2, the VQE framework determines the magnetic ground state of $CuSb_2O_6$ by mapping the anisotropic Heisenberg Hamiltonian, including MCA terms, into tensor-product Pauli operators. Spin orientations under an external magnetic field ($H$) are encoded in a parameterized Ansatz, optimized within the VQE framework, allowing explicit treatment of MCA's angular dependence.

For $CuSb_2O_6$, where spins form a nearly square lattice, the MCA energy can be expressed directly in the $a$, $b$, and $c$ directions, assuming invariant under spin inversion with respect to the $ab$ plane. This symmetry restricts the anisotropy to quadratic spin terms,

$$H_{\text{MCA}} = - \sum_{\alpha,\beta=a,b} \left(S_\alpha K_{\alpha\beta} S_\beta\right) + K_{cc} S_c^2,$$

where $K_{\alpha\beta}$ is a symmetric $2 \times 2$ matrix. Using the classical spin parameterization, $\boldsymbol{S} = (sin\theta cos\phi, sin\theta sin\phi, cos\theta)$, this yields

$$E_{\text{MCA}} = -K_{aa} sin^2\theta cos^2\phi - K_{bb} sin^2\theta sin^2\phi - K_{ab} sin^2\theta \, sin\, 2\phi + K_{cc} cos^2\theta.$$

In phenomenological models, $\boldsymbol{S}$ denotes a coarse-grained classical spin. For a quantum spin-1/2, however, $\boldsymbol{S}$ must be replaced by $\frac{\hbar}{2}\boldsymbol{\sigma}$, where Pauli operators $(X, Y, Z)$ correspond to the crystallographic axes $(a, b, c)$. Since the square of any Pauli operator is an identity operator, diagonal quadratic terms such as $S_\alpha^2$ ($\alpha = a, b, c$) reduce to constants, while off-diagonal terms vanish due to $\{X, Y\} = 0$. Consequently,

$$H_{\text{MCA}} = -K_{aa} - K_{bb} + K_{cc},$$

losing its dependence on spin orientation and thus failing to describe anisotropy. This limitation necessitates an alternative encoding capable of retaining quadratic spin terms.

To address this, we adopt a two-qubit encoding per spin site:

$$\mathcal{H}_{\text{MCA}} = -\sum_{n=1}^{4}(K_a X_n X_{n+4} + K_b Y_n Y_{n+4} - K_c Z_n Z_{n+4}),$$

where $X_n, Y_n, Z_n$ act on the $n$-th system qubit, and $n + 4$ denotes the paired ancilla. Here, $K_a \equiv K_{aa}$, $K_b \equiv K_{bb}$ and $K_c \equiv K_{cc}$ are the anisotropy constants along the $a$, $b$ and $c$ axes, respectively. In this mapping, $Z_n Z_{n+4}$ encodes the out-of-plane term $cos^2\theta_n$, while $X_n X_{n+4}$ and $Y_n Y_{n+4}$ represent the planar components $(sin\theta_n\, cos\phi_n)^2$ and $(sin\theta_n\, sin\phi_n)^2$. This construction preserves the full polar and azimuthal dependence of MCA within the Pauli-operator framework, allowing explicit representation in VQE simulations. As summarized in Fig. 2, this two-qubit mapping constitutes the central step of the VQE workflow, with further details provided in the Methods section.

**Energy competition and spin-flop transition under ultra-weak anisotropy**

The isothermal magnetization along the $b$ axis, $M_b$, calculated from the quantum Heisenberg frustrated-spin model with MCA (Fig. 3a), was fitted to experimental data at $T = 2$ K (Fig. 3b). To account for the $b$-axis chain alignment of $Cu^{2+}$ spins, the MCA constants were set smallest along $b$, intermediate along $a$, and largest along $c$ (see Methods). The low-$H_b$ slope was additionally fitted by assuming an 80% dominant twin domain population with spin chains along $b$. The corresponding results for the other orientations are shown in Supplementary Fig. S4. This fit yields $|K_b/J_1| = 0.00022$ % with $K_y = 0.00023$ meV. The simulated $M_b$ reproduces the experimental data well, validating the model for analyzing the $H_b$ dependence.

A distinct increase in $M_b$ appears at $H_{\text{flop}} = 1.2$ T, signaling a spin-flop transition. Despite the ultra-weak MCA, the relatively high $H_{\text{flop}}$ reflects the dominance of strong exchange interaction. The linear extrapolation of $M_b$ above $H_{\text{flop}}$ passes through the origin, a characteristic of a spin-flop transition, and agrees well with experiment (Fig. 3b). The schematics in Fig. 3a depict the spin reorientation, where spins rotate from the $b$ axis to the $a$ axis while maintaining ferromagnetic chain alignment along *b*. At higher $H_b$, only minimal canting occurs (~0.97° at $H_b = 3.5$ T, $M_b = 0.017$ μB).

Figures 3c and 3d show the $H_b$-dependent exchange, Zeeman, MCA, and total energies. At zero $H_b$, the large exchange energy (−38.306 meV) stabilizes the ferromagnetic chains, while the small MCA energy (−0.00023 meV) sets *b* as the easy axis. At $H_{\text{flop}}$, spins rotate toward the intermediate *a* axis, raising the MCA energy to +0.000012 meV and slightly perturbing the exchange energy. However, the Zeeman energy gain from slight canting outweighs both contributions, reducing the total energy and driving the transition. With further increase of $H_b$, the exchange energy continues to rise due to enhanced canting, but the Zeeman term dominates, producing an overall decrease in total energy.

To highlight the intrinsic balance, the small twin-domain contribution responsible for the low-$H_b$ slope was omitted, yielding flat low-$H_b$ energy curves. Direct comparison of the energy components clarifies how the spin-flop transition occurs despite the ultra-weak anisotropy.

**Emergence of a half-saturated magnetic phase under ultra-high magnetic fields**

As shown in Fig. 4a, beyond the spin-flop transition at $H_{\text{flop}} = 1.2$ T, $M_b$ increases linearly with $H_b$ as spins gradually cant toward the *b* axis. This evolution reflects the role of anisotropic NNN interactions: the stronger antiferromagnetic $J_2$ coupling between sites 1 and 3 resists canting and keeps these spins tilted more slowly, while the weaker $J_2'$ coupling allows sites 2 and 4 to align more readily with $H_b$.

Once $H_b$ exceeds $H_1 = 188.1$ T, spins on sites 2 and 4 saturate along the *b* axis, whereas spins on sites 1 and 3 continue to tilt gradually. This produces a partially saturated state in which only half the spins are fully aligned. With further increase of $H_b$, the slope of the $M_b$ decreases, reflecting the delayed alignment of the remaining spins, until full saturation is reached at $H_2 = 305.2$ T.

Figures 4b and 4c present the $H_b$-dependent energy landscape. Figure 4b shows the total energy together with exchange, Zeeman, and MCA contributions, while Fig. 4c resolves the exchange energy into individual paths. With increasing $H_b$, MCA energy decreases slightly as spins cant toward the easy *b* axis, while exchange energy rises due to perturbation of the spin-flop state. Anisotropic tilting modestly enhances $J_1$, whereas canting toward the *b* axis markedly reduces $J_1'$. Meanwhile, the NNN antiferromagnetic couplings grow substantially as spins deviate from antiparallel alignment: $J_2'$ saturates at $H_1$, while $J_2$ continues to increase through the half-saturated regime before leveling off at $H_2$. Nevertheless, the dominant Zeeman gain drives a continuous decrease of the total energy.

Although such ultra-high fields lie far beyond the reach of most pulsed-field experiments,[34,46,47] quantum simulations reveal this previously unrecognized half-saturated phase. These findings show that anisotropic NNN interactions not only generate frustration but also give rise to and stabilize new intermediate state.

**Angular torque signature of spin reorientation in CuSb$_2$O$_6$**

Despite the extremely small MCA in CuSb$_2$O$_6$, magnetic torque per unit volume, $\boldsymbol{\tau} = \boldsymbol{M} \times \boldsymbol{H}$, exhibits a pronounced angular dependence (Fig. 5), providing a sensitive probe of spin anisotropy. For these measurements, $H$ is rotated within the $bc$ plane, with the rotation angle $\alpha_H$ defined from the $b$ axis ($\alpha_H = 0°$ along the $b$ axis and $\alpha_H = 90°$ along the $c$ axis).

At $H$ = 1 T, $\tau$ is zero at $\alpha_H = 0°$, as the spins remain in a robust $b$-axis chain configuration. Rotating $H$ away from $b$ (e.g., $\alpha_H = 45°$) introduces the $c$-axis component, canting the net moment toward $c$ ($M_c = 0.00403$ μ$_B$) and producing a finite negative $\tau$ (Fig. 5a). This canting increases the MCA energy by misaligning spins from the easy axis, while a small $b$-axis component of $H$ deflects the moment slightly toward $b$ ($M_b = 0.000831$ μ$_B$). At $\alpha_H = 90°$, the moment aligns fully along $c$, restoring zero $\tau$. Further rotation reverses the relative orientation between $M$ and $H$, yielding positive $\tau$. Overall, $\tau(\alpha_H)$ follows a near-sinusoidal variation.

At $H$ = 1.5 T, above the $H_{flop}$, the spins reorient into a spin-flop state aligned mainly along $a$, with a small $b$-axis moment parallel to $H$, again giving $\tau$ = 0 at $\alpha_H = 0°$ (Fig. 5b). Zero $\tau$ persists as long as the $b$-axis $H$ component exceeds $H_{flop}$. While simulations maintain zero $\tau$ up to the critical angle, measured data show a slight negative $\tau$ early in rotation, possibly due to minority twin domains responding earlier to the $c$-axis $H$. Once the $b$-axis $H$ falls below $H_{flop}$, the spin-flop state collapses, and the net moment shifts mainly toward $c$, causing a sharp $\tau$ drop. Beyond $\alpha_H = 90°$, the sign of $\tau$ reverses as $M$ changes its relative orientation to $H$. This angular profile directly traces the collapse and recovery of the spin-flop state as the $b$-axis $H$ component crosses $H_{flop}$. $\tau$ analysis thus serves as a complementary and highly sensitive probe of spin reorientation, revealing anisotropy and domain effects that are less accessible through magnetization alone.

## Discussion

In summary, our study demonstrates that quantum simulation can accurately resolve MCA in a real spin-1/2 antiferromagnet, revealing both the subtle spin-flop transition and an unexpected half-saturated phase under ultra-high fields. Beyond reproducing experimental signatures such as torque anisotropy, the results show how anisotropic NNN couplings shape unconventional magnetic states.

This framework offers a resource-efficient route to include MCA—a crucial yet often overlooked ingredient in spin models—within realistic quantum simulations. Validated through Qiskit statevector simulations, it bridges the gap between theoretical models and experimental observations. At the same time, Qiskit shot-based simulations reveal the current limitations of noisy devices, where statistical noise obscures the subtle MCA signal, pointing to the need for processors with longer coherence and lower error rates.

Looking forward, fault-tolerant quantum platforms with logical qubits could reliably resolve sub-µeV anisotropies, extend simulations to larger lattices and dynamical regimes, and provide predictive insights into both fundamental magnetism and the design of spintronic materials. Such advances promise to expand the scope of quantum simulation in materials research, opening pathways not accessible to classical methods or today's hardware.

## Methods

### Sample preparation and physical property characterization

Single crystals of $CuSb_2O_6$ were grown by the chemical vapor transport method, using $TeCl_4$ as the transport agent.[36,48] Temperature-dependent magnetic susceptibility and field-dependent magnetization were measured with a vibrating sample magnetometer (VSM) integrated into a Physical Property Measurement System (PPMS, Quantum Design, Inc., USA). Magnetic torque measurements were carried out in the PPMS using a calibrated cantilever chip (P109A, Quantum Design, Inc., USA) mounted on a single-axis rotator.

### Quantum Hamiltonian with magnetocrystalline anisotropy (MCA)

The $Cu^{2+}$ ions in $CuSb_2O_6$ form a nearly square lattice with four magnetic sites ($n =$ 1, 2, 3, 4), as shown in Fig. 1b. The total Hamiltonian incorporates anisotropic nearest-neighbor and next-nearest-neighbor exchange couplings, Zeeman interaction, and MCA terms:

$$\mathcal{H} = \frac{J_1 S^2}{2} \sum_{\substack{n=1 \\ n \text{ odd}}}^{4} (X_n X_{n+1} + Y_n Y_{n+1} + Z_n Z_{n+1}) + \frac{J_1' S^2}{2} \sum_{\substack{n=1 \\ n \text{ even}}}^{4} (X_n X_{n+1} + Y_n Y_{n+1} + Z_n Z_{n+1})$$

$$+ J_2 S^2 (X_1 X_3 + Y_1 Y_3 + Z_1 Z_3) + J_2' S^2 (X_2 X_4 + Y_2 Y_4 + Z_2 Z_4) - g\mu_B H_b S \sum_n Y_n$$

$$- \sum_{n=1}^{4} (K_a X_n X_{n+4} + K_b Y_n Y_{n+4} - K_c Z_n Z_{n+4}),$$

where $X_n$, $Y_n$, and $Z_n$ are Pauli matrices acting on the $n$-th qubit, $n+4$ denotes its ancilla qubit, $S$ is the magnitude of $Cu^{2+}$ spin, $g = 2$ is the Landé-$g$-factor, $\mu_B$ is the Bohr magneton. The fifth term is the Zeeman interaction under $H_b$. When the magnetic field lies in the $bc$ plane and is rotated by an angle $\alpha_H$ for magnetic torque measurements, the Zeeman term becomes

$$H_{\text{Zeeman}} = -g\mu_B H_b S \left( \cos \alpha_H \sum_{n=1}^{4} Y_n + \sin \alpha_H \sum_{n=1}^{4} Z_n \right).$$

The last term introduces the MCA contribution, discussed in detail below.

**Two-qubit encoding for MCA**

To encode MCA explicitly, we employ a two-qubit representation per spin site, consisting of a system qubit and an ancilla. The full Hamiltonian consists of the system Hamiltonian and the coupling to the ancilla qubits:

$$\mathcal{H} = H_{\text{sys}} - \sum_{n=1}^{4} (K_a X_n X_{n+4} + K_b Y_n Y_{n+4} - K_c Z_n Z_{n+4}).$$

The coupling terms emulate the MCA. The ground state is described by the variational Ansatz $\Psi(\{\theta_n, \phi_n\})$ ($n = 1, \cdots, 8$). The corresponding expectation value is

$$E(\{\theta_n, \phi_n\}) = E_{\text{sys}}(\{\theta_n, \phi_n\}) - \sum_{n=1}^{4} (K_a \langle X_n \rangle \langle X_{n+4} \rangle + K_b \langle Y_n \rangle \langle Y_{n+4} \rangle - K_c \langle Z_n \rangle \langle Z_{n+4} \rangle).$$

MCA is implemented by imposing the condition $\langle \sigma_n \rangle = \langle \sigma_{n+4} \rangle$, achieved by restricting the variational parameters such that $\{\theta_n, \phi_n\} = \{\theta_{n+4}, \phi_{n+4}\}$. This effectively generates the MCA contribution in the VQE analysis. Importantly unlike system qubits, the ancilla qubits are not used for Hamiltonian optimization but serve solely to enforce this constraint.

**Variational Ansatz and optimization**

Each spin orientation $(\theta_n, \phi_n)$ is encoded through single-qubit rotations $R_Y(\theta_n)$ and $R_Z(\phi_n)$. The full variational state, $\Psi(\{\theta_n, \phi_n\})$ is an eight-qubit tensor product under the two-qubit mapping per spin site. The variational quantum eigensolver minimizes the expectation value $\langle\hat{\mathcal{H}}\rangle(\{\theta_n, \phi_n\}) = \langle\Psi(\{\theta_n, \phi_n\})|\hat{\mathcal{H}}|\Psi(\{\theta_n, \phi_n\})\rangle$ using the Nelder–Mead optimizer. All calculations were performed with the Qiskit statevector simulator, which provides an ideal, noise-free emulation of the quantum algorithm's operations. Fitting the theoretically calculated $M_b$ to experimental data at $T = 2$ K produces the following parameters: $J_1 = -104.30$ meV, $J_1' = -103.13$ meV, $J_2 = 87.18$ meV, $J_2' = 64.87$ meV, $K_a = 0.000012$ meV, $K_b = 0.00023$ meV, and $K_c = 0.00010$ meV.

**References**

1. Lloyd, S. Universal quantum simulators. *Science* **273**, 1073-1078 (1996).
2. Feynman, R. P. in *Feynman and computation* 133-153 (cRc Press, 2018).
3. Georgescu, I. M., Ashhab, S. & Nori, F. Quantum simulation. *Reviews of Modern Physics* **86**, 153-185 (2014).
4. Preskill, J. Quantum computing in the NISQ era and beyond. *Quantum* **2**, 79 (2018).
5. Daley, A. J. *et al.* Practical quantum advantage in quantum simulation. *Nature* **607**, 667-676 (2022).
6. Alexeev, Y. *et al.* Quantum-centric supercomputing for materials science: A perspective on challenges and future directions. *Future Generation Computer Systems* **160**, 666-710 (2024).
7. Peruzzo, A. *et al.* A variational eigenvalue solver on a photonic quantum processor. *Nature communications* **5**, 4213 (2014).
8. McClean, J. R., Romero, J., Babbush, R. & Aspuru-Guzik, A. The theory of variational hybrid quantum-classical algorithms. *New Journal of Physics* **18**, 023023 (2016).
9. Kandala, A. *et al.* Hardware-efficient variational quantum eigensolver for small molecules and quantum magnets. *Nature* **549**, 242-246 (2017).
10. Tilly, J. *et al.* The variational quantum eigensolver: a review of methods and best practices. *Physics Reports* **986**, 1-128 (2022).
11. Romero, J. *et al.* Strategies for quantum computing molecular energies using the unitary coupled cluster ansatz. *Quantum Science and Technology* **4**, 014008 (2018).
12. Hong, C.-L. *et al.* Accurate and efficient quantum computations of molecular properties

using daubechies wavelet molecular orbitals: A benchmark study against experimental data. *PRX Quantum* **3**, 020360 (2022).

13. O'Malley, P. J. *et al.* Scalable quantum simulation of molecular energies. *Physical Review X* **6**, 031007 (2016).

14. Hempel, C. *et al.* Quantum chemistry calculations on a trapped-ion quantum simulator. *Physical Review X* **8**, 031022 (2018).

15. Zhang, Y. *et al.* Variational quantum eigensolver with reduced circuit complexity. *npj Quantum Information* **8**, 96 (2022).

16. Salathé, Y. *et al.* Digital quantum simulation of spin models with circuit quantum electrodynamics. *Physical Review X* **5**, 021027 (2015).

17. Heras, U. L. *et al.* Digital quantum simulation of spin systems in superconducting circuits. *Physical Review Letters* **112**, 200501 (2014).

18. Porras, D. & Cirac, J. I. Effective quantum spin systems with trapped ions. *Physical Review Letters* **92**, 207901 (2004).

19. Tennant, D. A., Cowley, R. A., Nagler, S. E. & Tsvelik, A. M. Measurement of the spin-excitation continuum in one-dimensional KCuF 3 using neutron scattering. *Physical Review B* **52**, 13368 (1995).

20. Hase, M., Terasaki, I. & Uchinokura, K. Observation of the spin-Peierls transition in linear $Cu^{2+}$(spin-1/2) chains in an inorganic compound $CuGeO_3$. *Physical Review Letters* **70**, 3651 (1993).

21. Schlappa, J. *et al.* Probing multi-spinon excitations outside of the two-spinon continuum in the antiferromagnetic spin chain cuprate $Sr_2CuO_3$. *Nature communications* **9**, 5394 (2018).

22. Azuma, M., Hiroi, Z., Takano, M., Ishida, K. & Kitaoka, Y. Observation of a Spin Gap in $SrCu_2O_3$ Comprising Spin-½ Quasi-1D Two-Leg Ladders. *Physical Review Letters* **73**, 3463 (1994).

23. Klanjšek, M. *et al.* Controlling Luttinger liquid physics in spin ladders under a magnetic field. *Physical Review Letters* **101**, 137207 (2008).

24. Nagata, T. *et al.* Pressure-Induced Dimensional Crossover and Superconductivity in the Hole-Doped Two-Leg Ladder Compound $Sr_{14-x}Ca_x Cu_{24}O_{41}$. *Physical Review Letters* **81**, 1090 (1998).

25. Coldea, R. *et al.* Spin waves and electronic interactions in $La_2CuO_4$. *Physical Review Letters* **86**, 5377 (2001).

26. Han, T.-H. *et al.* Fractionalized excitations in the spin-liquid state of a kagome-lattice antiferromagnet. *Nature* **492**, 406-410 (2012).
27. Ma, J. *et al.* Static and dynamical properties of the spin-1/2 equilateral triangular-lattice antiferromagnet $Ba_3CoSb_2O_9$. *Physical Review Letters* **116**, 087201 (2016).
28. Zhu, J.-X. *et al.* LDA+ DMFT approach to magnetocrystalline anisotropy of strong magnets. *Physical Review X* **4**, 021027 (2014).
29. Bruno, P. Tight-binding approach to the orbital magnetic moment and magnetocrystalline anisotropy of transition-metal monolayers. *Physical Review B* **39**, 865 (1989).
30. Bogdanov, A., Zhuravlev, A. & Rößler, U. Spin-flop transition in uniaxial antiferromagnets: Magnetic phases, reorientation effects, and multidomain states. *Physical Review B—Condensed Matter and Materials Physics* **75**, 094425 (2007).
31. Basnet, R., Wegner, A., Pandey, K., Storment, S. & Hu, J. Highly sensitive spin-flop transition in antiferromagnetic van der Waals material $MPS_3$(M= Ni and Mn). *Physical Review Materials* **5**, 064413 (2021).
32. Seo, J. Y. *et al.* Probing the weak limit of magnetocrystalline anisotropy through a spin–flop transition in the van der Waals antiferromagnet $CrPS_4$. *NPG Asia Materials* **16**, 39 (2024).
33. Povarov, K. Y., Smirnov, A. & Landee, C. Switching of anisotropy and phase diagram of the Heisenberg square-lattice S= 1/2 antiferromagnet $Cu(pz)_2(ClO_4)_2$. *Physical Review B—Condensed Matter and Materials Physics* **87**, 214402 (2013).
34. Matsuda, Y. H. *et al.* Magnetization of $SrCu_2(BO_3)_2$ in ultrahigh magnetic fields up to 118 T. *Physical Review Letters* **111**, 137204 (2013).
35. Zheludev, A. *et al.* Field-induced commensurate-incommensurate phase transition in a Dzyaloshinskii-Moriya spiral antiferromagnet. *Physical Review Letters* **78**, 4857 (1997).
36. Heinrich, M. *et al.* Structural and magnetic properties of $CuSb_2O_6$ probed by ESR. *Physical Review B* **67**, 224418 (2003).
37. Rebello, A., Smith, M., Neumeier, J., White, B. & Yu, Y.-K. Transition from one-dimensional antiferromagnetism to three-dimensional antiferromagnetic order in single-crystalline $CuSb_2O_6$. *Physical Review B—Condensed Matter and Materials Physics* **87**, 224427 (2013).
38. Herak, M., Žilić, D., Matković Čalogović, D. & Berger, H. Torque magnetometry study

of magnetically ordered state and spin reorientation in the quasi-one-dimensional S= 1 2 Heisenberg antiferromagnet $CuSb_2O_6$. *Physical Review B* **91**, 174436 (2015).

39. Giere, E.-O., Brahimi, A., Deiseroth, H. & Reinen, D. The Geometry and Electronic Structure of the $Cu^{2+}$ Polyhedra in Trirutile-Type Compounds $Zn(Mg)_{1-x}Cu_xSb_2O_6$ and the Dimorphism of $CuSb_2O_6$: A Solid State and EPR Study. *Journal of Solid State Chemistry* **131**, 263-274 (1997).

40. Nakua, A., Yun, H., Reimers, J., Greedan, J. & Stager, C. Crystal structure, short range and long range magnetic ordering in $CuSb_2O_6$. *Journal of solid state chemistry* **91**, 105-112 (1991).

41. Kato, M. *et al.* Magnetic structure of $CuSb_2O_6$. *Journal of the Physical Society of Japan* **71**, 187-189 (2002).

42. Kasinathan, D., Koepernik, K. & Rosner, H. Quasi-one-dimensional magnetism driven by unusual orbital ordering in $CuSb_2O_6$. *Physical Review Letters* **100**, 237202 (2008).

43. Maimone, D., Christian, A., Neumeier, J. & Granado, E. Lattice dynamics of $ASb_2O_6$ (A= Cu, Co) with trirutile structure. *Physical Review B* **97**, 104304 (2018).

44. Goodenough, J. B. Theory of the role of covalence in the perovskite-type manganites [La, M(II)]$MnO_3$. *Physical Review* **100**, 564 (1955).

45. Kanamori, J. Superexchange interaction and symmetry properties of electron orbitals. *Journal of Physics and Chemistry of Solids* **10**, 87-98 (1959).

46. Jaime, M. *et al.* Magnetostriction and magnetic texture to 100.75 Tesla in frustrated $SrCu_2(BO_3)_2$. *Proceedings of the National Academy of Sciences* **109**, 12404-12407 (2012).

47. Gen, M. *et al.* Higher magnetic-field generation by a mass-loaded single-turn coil. *Review of Scientific Instruments* **92** (2021).

48. Prokofiev, A., Ritter, F., Assmus, W., Gibson, B. & Kremer, R. Crystal growth and characterization of the magnetic properties of $CuSb_2O_6$. *Journal of crystal growth* **247**, 457-466 (2003).

**Acknowledgements**

This work was supported by the National Research Foundation of Korea (NRF) through grants RS-2021-NR058241, NRF-2022M3K2A108385813, RS-2023-00257561, RS-2025-09192968, , RS-2025-25443863, and RS-2025-16070108.

**Author contributions**

K. M., N. L. and Y. J. C. initiated and supervised the project. J. Y. S. synthesized the single crystals. J. Y. S., J. M. H., and N. L. performed measurements of physical properties. K. W. J., S. L., H. J. R., J. S. B. and K. M. carried out the quantum computations. K. W. J., J. Y. S., K. M., N. L., and Y. J. C. analyzed the data and prepared the manuscript. All authors have read and approved the final version of the manuscript.

**Competing interests**

The authors declare no competing interests.

**Additional information**

**Supplementary information** The online version contains supplementary material

**Correspondence** and requests for materials should be addressed to Kyungsun Moon, Nara Lee or Young Jai Choi.

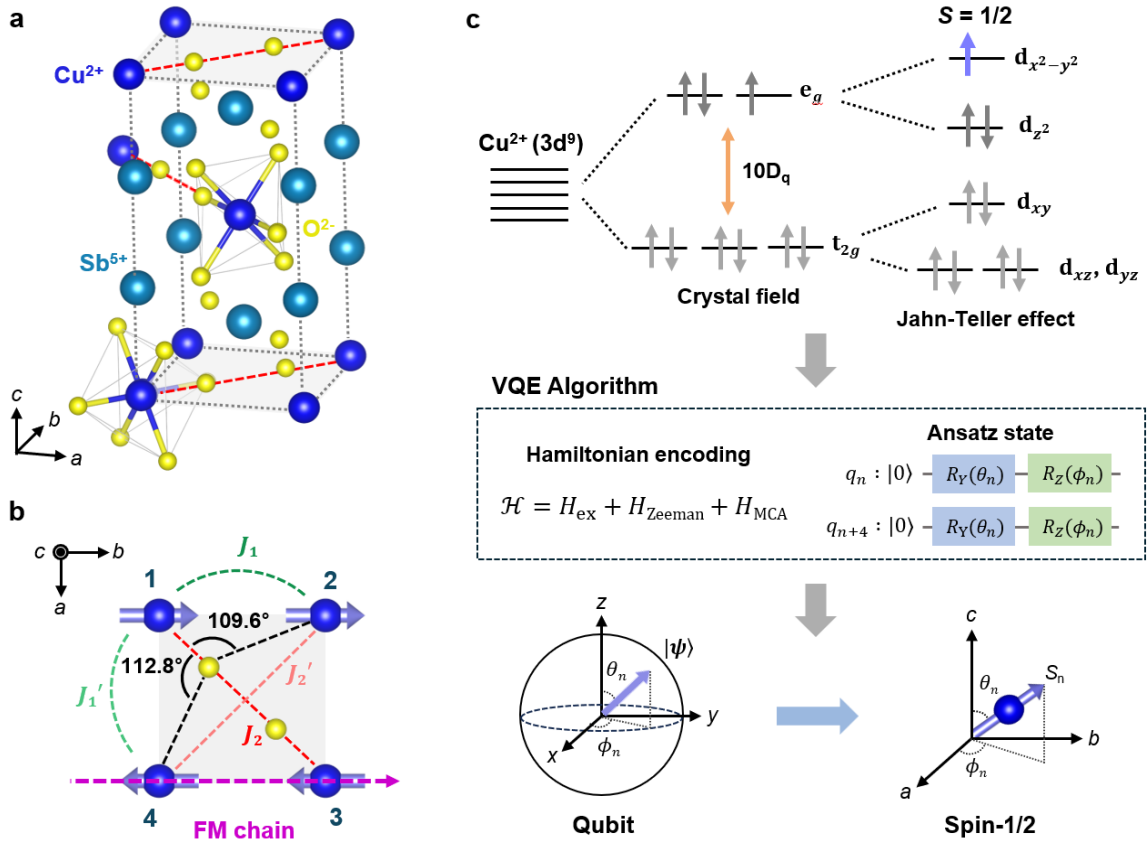

**Fig. 1 Crystallographic structure, exchange network, and quantum-simulation framework for CuSb$_2$O$_6$. a** Monoclinic crystal structure in the low-$T$ phase. Blue, teal, and yellow spheres represent Cu$^{2+}$, Sb$^{5+}$, and O$^{2-}$ ions, respectively. **b** Spin configuration on the nearly square lattice within the *ab* plane. Cu$^{2+}$ ions form magnetic sublattices via oxygen-mediated superexchange paths (dashed curves and lines). Light purple arrows on Cu$^{2+}$ sites indicate spin orientations, showing ferromagnetic (FM) chain formation along the *b* axis (magenta dashed arrow), with alternating antiferromagnetic interchain alignment. Sites 1 to 4 are labeled clockwise from the top left. $J_1$ and $J_1'$ are the dominant nearest-neighbor (NN) exchanges along the *b* and *a* axes, respectively, with bond angles indicated. $J_2$ and $J_2'$ represent next-nearest-neighbor (NNN) superexchange couplings along the diagonals. **c** Quantum simulation framework for Cu$^{2+}$ spins. The Cu$^{2+}$ ion (3d$^9$) undergoes crystal field splitting into t$_{2g}$ and e$_g$ orbitals, followed by a static Jahn–Teller distortion that lifts the e$_g$ degeneracy, and stabilizes a single unpaired electron in the d$_{x^2-y^2}$ orbital. This leads to an effective spin-1/2, allowing a direct one-to-one mapping to a qubit. The variational quantum eigensolver (VQE) algorithm estimates the ground state energy of $\mathcal{H} = H_{\text{ex}} + H_{\text{Zeeman}} + H_{\text{MCA}}$, with each term

representing exchange, Zeeman and magnetocrystalline ansitoropy (MCA) contributions. The variational Ansatz uses parameterized $R_Y(\theta_n)$ and $R_Z(\phi_n)$ gates on each site, with the resulting qubit state reflecting local spin orientation and demonstrating the equivalence between spin-1/2 particles and qubits for simulating complex magnetism.

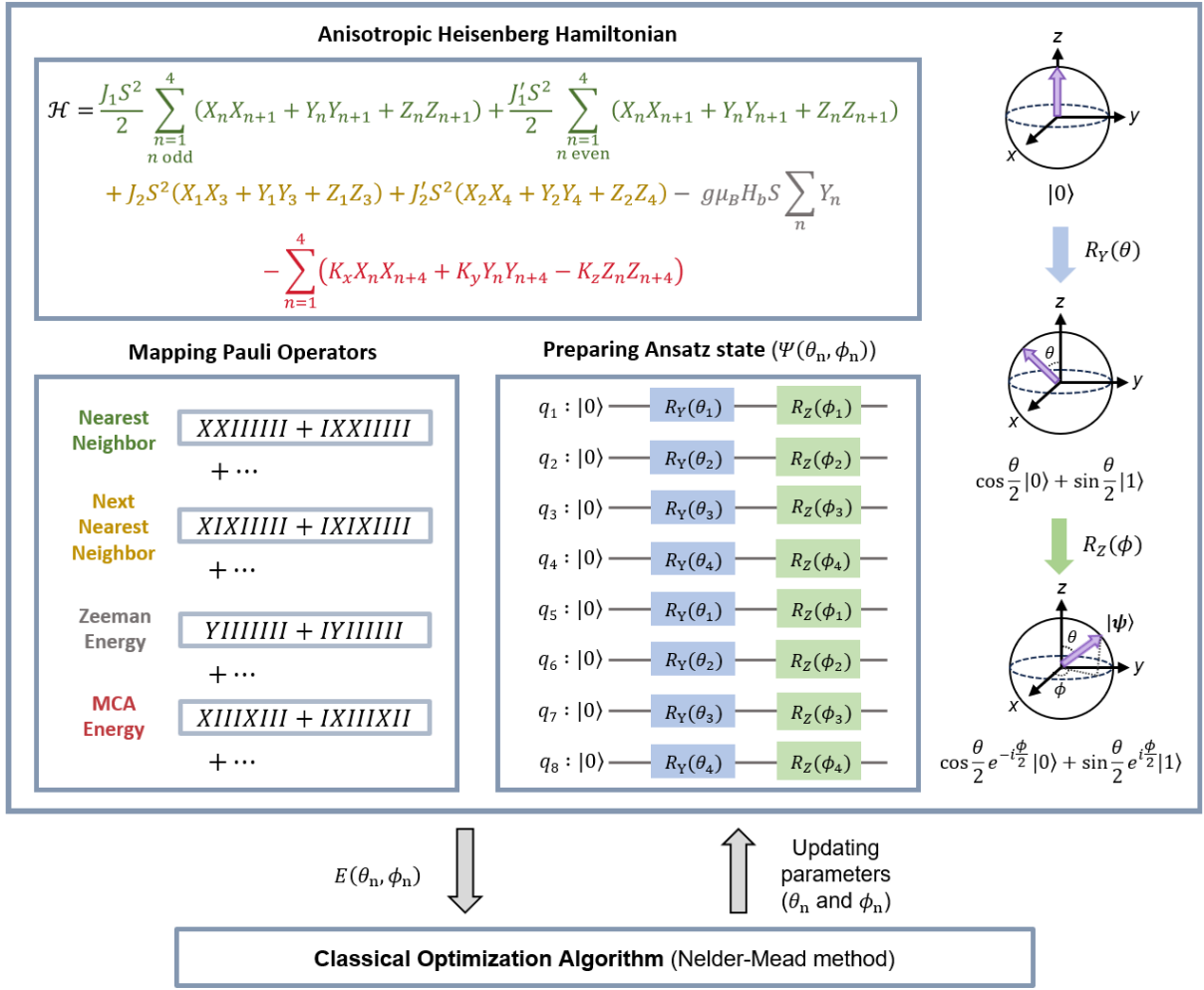

**Fig. 2 VQE framework for CuSb$_2$O$_6$.** Top box: Anisotropic Heisenberg Hamiltonian including NN ($J_1, J_1'$) and NNN ($J_2, J_2'$) exchange interactions, Zeeman interaction under $H_b$, and MCA terms ($K_a, K_b, K_c$). Bottom left: Encoding of Hamiltonian terms into Qiskit's eight-qubit representation. Each Cu$^{2+}$ spin site is mapped to two qubits, as described in the main text, so that quadratic spin projections required for MCA are preserved through tensor-product Pauli strings. Bottom right and rightmost schematic: Variational Ansatz preparation using $R_Y(\theta_n)$ and $R_Z(\phi_n)$ gates with $\theta_n$ and $\phi_n$ representing the polar and azimuthal spin angles. Workflow arrow: Iterative VQE cycle of state preparation, Hamiltonian expectation measurement, and Nelder–Mead parameter update, converging to the optimized spin configuration and enabling simulation of the spin-flop transition.

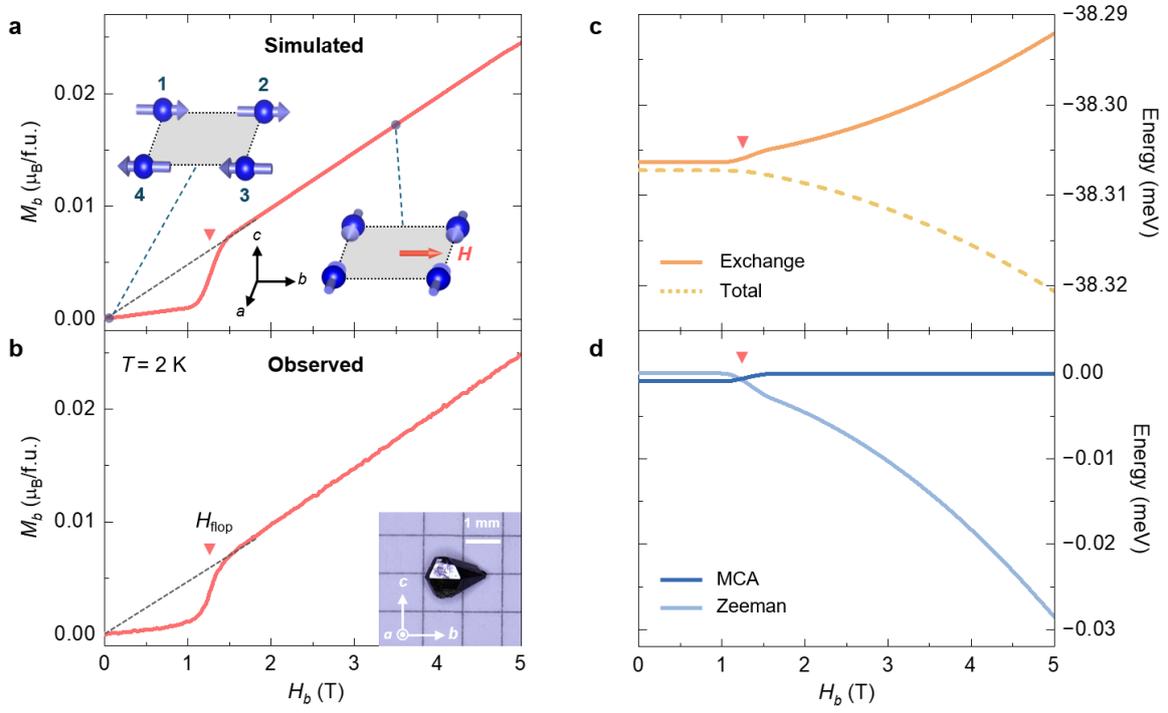

**Fig. 3 Magnetic energy balance underlying the spin-flop transition in CuSb$_2$O$_6$. a** Simulated magnetization along the $b$ axis ($M_b$) from the quantum Heisenberg model with MCA, showing a spin-flop transition at $H_{\text{flop}} = 1.2$ T. Insets illustrate spin configurations below and above $H_{\text{flop}}$, where spins rotate by 90° from the $b$ axis to the $a$ axis while retaining ferromagnetic chain alignment along the $b$ axis. **b** Experimental $M_b$ at $T = 2$ K for $H_b$. Inset shows optical image of a CuSb$_2$O$_6$ single crystal. **c** $H_b$-dependent exchange and total magnetic energies, showing the balance shift at $H_{\text{flop}}$. **d** MCA and Zeeman energy contributions, highlighting the Zeeman gain that drives the transition. Red triangles mark $H_{\text{flop}}$.

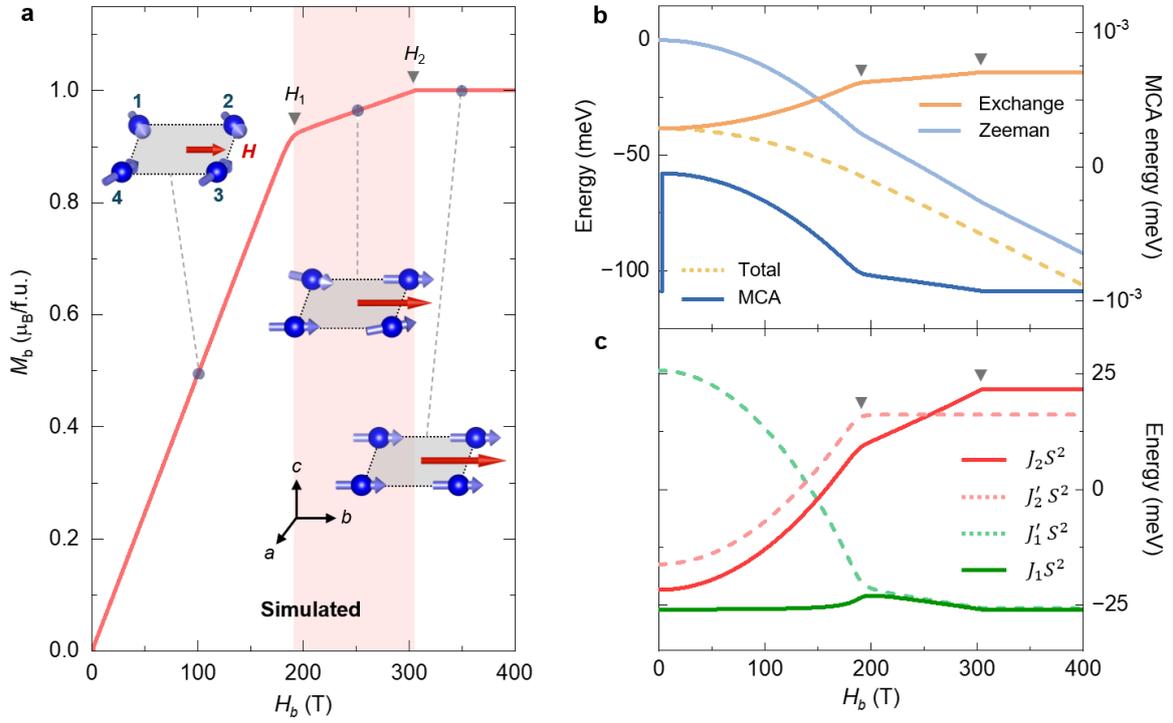

**Fig. 4 Half-saturated magnetic phase under ultra-high magnetic fields. a** Simulated $M_b$ up to 400 T, revealing the emergence of a half-saturated phase prior to full saturation. The red shaded region highlights the half-saturated regime. Insets depict spin configurations in the spin-flop, half-saturated and fully saturated states. **b** $H_b$-dependent energy balance showing exchange, Zeeman, MCA and total magnetic contributions in the high-$H_b$ regime. Because of its tiny scale, the MCA energy is plotted on the right $y$ axis. **c** Decomposition of the exchange energy into individual contributions. $H_b$-dependent asymmetry of $J_2$ and $J_2'$ stabilizes the half-saturated phase.

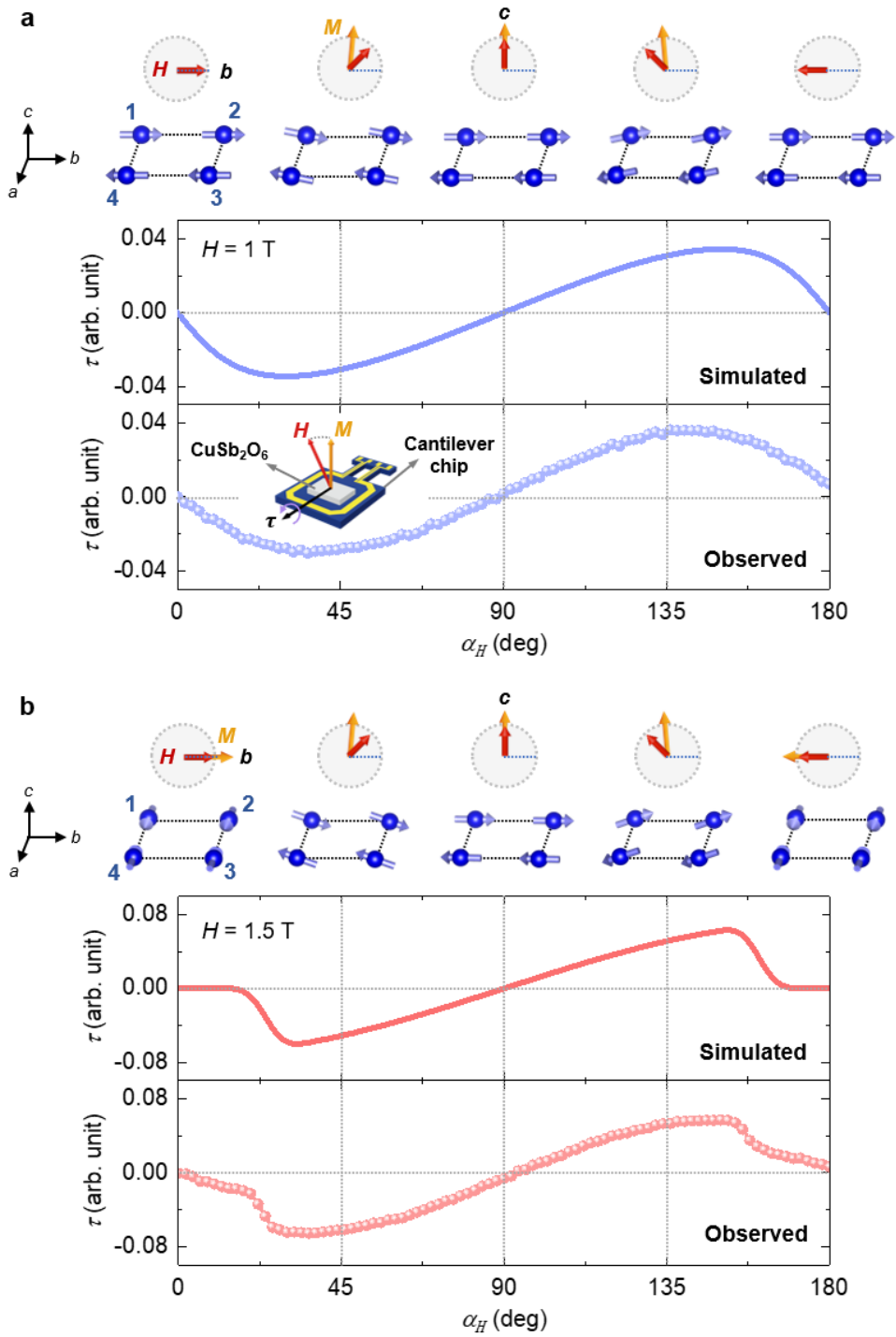

**Fig. 5 Angular dependence of magnetic torque across the spin-flop transition. a** Observed (bottom) and simulated (top) torque $\tau(\alpha_H)$ for $H = 1$ T with $H$ rotated in the $bc$ plane. Representative spin configurations below $H_{\text{flop}}$ illustrate the canting behavior. $\tau$ varies nearly sinusoidally, consistent with weak anisotropy. A schematic of the cantilever-based $\tau$ measurement is also shown. **b** Observed (bottom) and simulated (top) $\tau(\alpha_H)$ for $H = 1.5$ T with

$H$ rotated in the $bc$ plane. Insets show spin reorientation above $H_{\text{flop}}$. At this field, $\tau$ traces the collapse and recovery of the spin-flop state as the $b$-axis field component crosses $H_{\text{flop}}$, serving as a sensitive probe of spin reorientation under ultra-weak anisotropy.

Supplementary Information

# Quantum simulation approach to ultra-weak magnetic anisotropy in a frustrated spin-1/2 antiferromagnet

Ki Won Jeong[*], Jae Yeon Seo[*], Sunghyun Lim, Jae Min Hong, Hyeon Jun Ryu, Jongseok Byeon, Kyungsun Moon, Nara Lee, and Young Jai Choi

Department of Physics, Yonsei University, Seoul 03722, Korea

*These authors contributed equally to this work.
Correspondence: Kyungsun Moon (kmoon@yonsei.ac.kr), Nara Lee (eland@yonsei.ac.kr) or Young Jai Choi (phylove@yonsei.ac.kr)

## S1. Crystallographic twin domains in monoclinic $CuSb_2O_6$

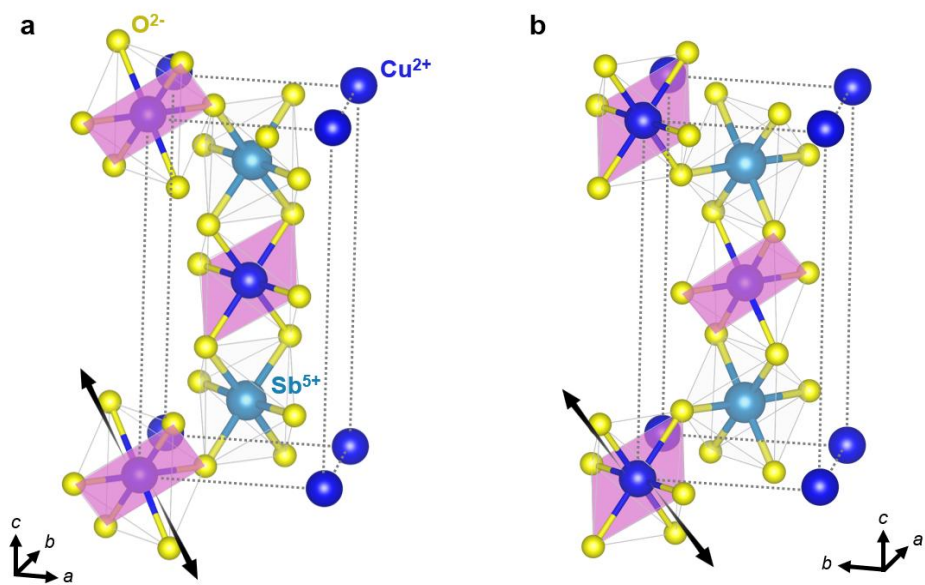

**Fig. S1 Crystal structure of monoclinic $CuSb_2O_6$ illustrating the two twin variants. a** and **b** show symmetry-related orientations of the elongated $CuO_6$ octahedra in the low-temperature (*T*) monoclinic phase. The Jahn-Teller distortion produces elongation axes tilted by ~12° from the normal to the equatorial plane, giving rise to two distinct twin domains. Equatorial planes are highlighted in pink, and the elongation directions are indicated by black arrows. Yellow,

blue, and teal spheres represent $O^{2-}$, $Cu^{2+}$, and $Sb^{5+}$ ions, respectively.

At 380 K, $CuSb_2O_6$ undergoes a structural phase transition from the high-$T$ tetragonal ($P4_2/nmn$) to the low-$T$ monoclinic ($P2_1/n$) structure, driven by a cooperative Jahn-Teller effect.[1,2] In the monoclinic phase, the four $O^{2-}$ ligands in the equatorial plane of each $CuO_6$ octahedron becomes inequivalent, resulting in a static elongation along one of two symmetry-allowed directions.[3,4] The elongation axis, tilted by ~12° from the normal to the equatorial plane, yields two energetically equivalent distortion variants.[1,2] These variants coexist within a single crystal, forming crystallographic twins with different orientations of the elongated $CuO_6$ octahedra (Fig. S1). Above 380 K, thermal fluctuations dynamically average the two distortions, restoring the tetragonal geometry via the dynamic Jahn–Teller effect.[5]

## S2. Temperature-dependent anisotropic magnetic susceptibility

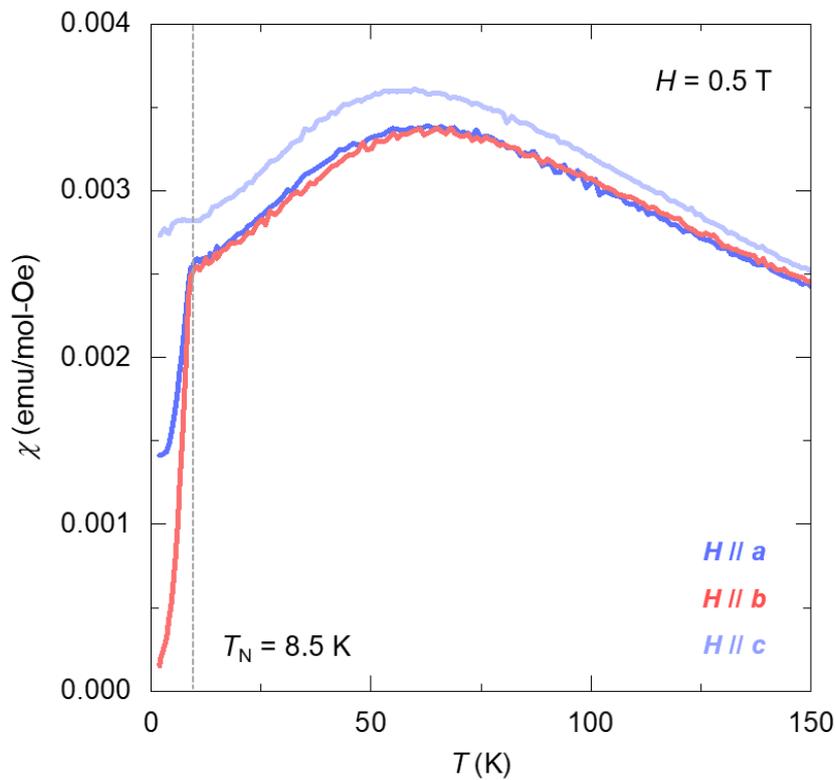

**Fig. S2 Magnetic susceptibility.** $T$-dependence of magnetic susceptibility for $H//a$, $H//b$ and

$H//c$, measured upon warming at $H = 0.5$ T. The vertical dotted gray line marks the Néel $T$ ($T_N = 8.5$ K).

The ultra-weak yet discernible magnetic anisotropy of $CuSb_2O_6$ is reflected in its $T$-dependent susceptibility, $\chi = M/H$. Because the anisotropy energy is much smaller than the dominant exchange interactions, the overall $\chi$ remains suppressed across the measured $T$ range, showing only modest field-induced responses. Nevertheless, distinct directional differences appear among the $a$, $b$, and $c$ axes. Measurements were performed at $H = 0.5$ T upon warming after zero-field cooling, with $H$ applied along each crystallographic axis (Fig. S2).

A sharp drop in $\chi$ for $H//b$ at $T_N = 8.5$ K signals the antiparallel ordering of ferromagnetic spin chains along the $b$ axis.[3] In contrast, the weaker reduction observed for $H//a$ indicates partial twin formation along the $a$ axis, consistent with earlier reports.[1,2] Additionally, the broad maximum near 60 K reflects the characteristic behavior of spin-1/2 chains described by the Bonner–Fisher model, which predicts a wide susceptibility peak in the intermediate-$T$ regime.[6]

## S3. Competing spin configurations in a square lattice

For an ideal square lattice with equal nearest-neighbor exchange interactions ($J_1 = J_1'$), the antiferromagnetic next-nearest-neighbor coupling ($J_2 > 0$) introduces magnetic frustration, yielding two degenerate spin-chain configurations. These states consist of ferromagnetic spin chains oriented along either the $a$ or $b$ axis, coupled antiferromagnetically between adjacent chains (Fig. S3).

In the real $CuSb_2O_6$ structure, this degeneracy is lifted by subtle structural anisotropy. Although the $b$-axis lattice parameter is only slightly larger than $a$, the $Cu^{2+}$–$O^{2-}$–$Cu^{2+}$ bond angle along $b$ (109.6°) is closer to the ideal 90° superexchange geometry than along $a$ (112.8°).[1,3] According to the Goodenough–Kanamori rules,[7,8] this favors stronger ferromagnetic exchange along $b$

($|J_1| > |J_1'|$), thereby stabilizing spin-chain alignment along the $b$ axis as the ground-state (Fig. S3b).

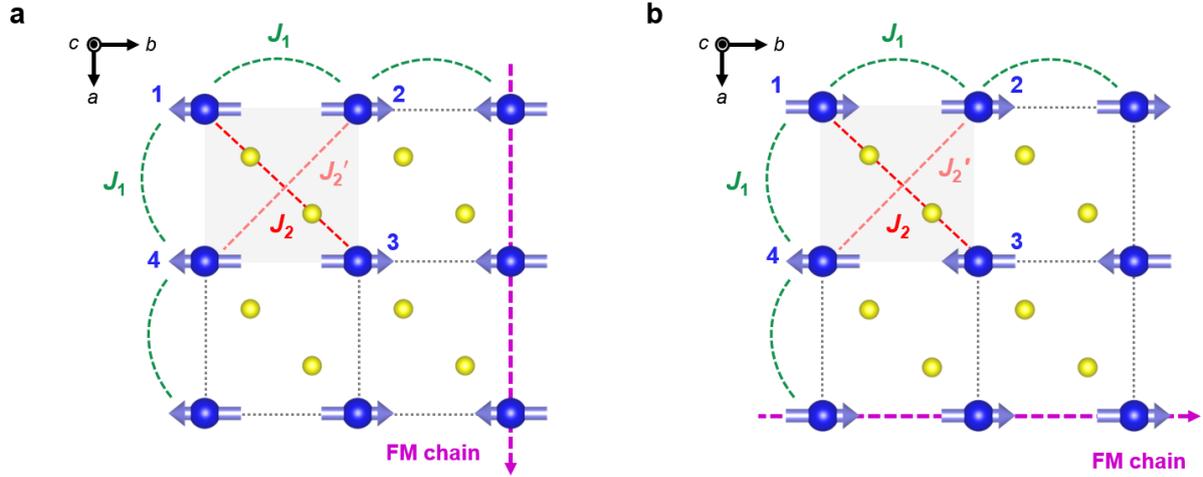

**Fig. S3 Lifting of spin-chain degeneracy in CuSb$_2$O$_6$.** Two degenerate configurations arise in an ideal square lattice when $J_1 = J_1'$. **a** Ferromagnetic (FM) chains aligned along the $a$ axis with antiferromagnetic coupling between chains. **b** Equivalent arrangement with chains aligned along the $b$ axis. In real CuSb$_2$O$_6$, this degeneracy is lifted by subtle anisotropy in bond angles and lattice parameters, stabilizing spin alignment along the $b$ axis.

**S4. Effect of crystallographic twins on magnetization along $a$ and $c$ axes.**

Isothermal magnetization along the $a$ and $c$ directions ($M_a$ and $M_c$) at $T$ = 2 K was analyzed using the same frustrated Heisenberg model, with MCA parameters determined from fitting the $b$-axis data and further refined by comparison with the $a$- and $c$-axis results (Fig. 3 in the main manuscript and Fig. S4). In the monoclinic low-$T$ phase, CuSb$_2$O$_6$ forms crystallographic twins with Jahn-Teller–elongated CuO$_6$ octahedra oriented along two symmetry-allowed directions (Fig. S1). The dominant twin preserves spin chains along the $b$ axis, while partial formation of the minor twin rotates ~20% of chains toward the $a$ axis. This twin fraction, estimated from the low-field slope of $M_b$ (Fig. 3 in the main manuscript), was incorporated into the simulations shown in Fig. S4.

For $M_a$, the 20% twin contribution produces a weak spin-flop feature superimposed on the otherwise linear field response, whereas $M_c$ shows no spin-flop anomaly. The experimentally observed spin-flop along the *a* axis is somewhat stronger than predicted, likely reflecting local inhomogeneities in twin distribution or microstructural effects (e.g., strain or domain boundaries) that enhance anisotropy beyond the uniform twin ratio assumed in the model.

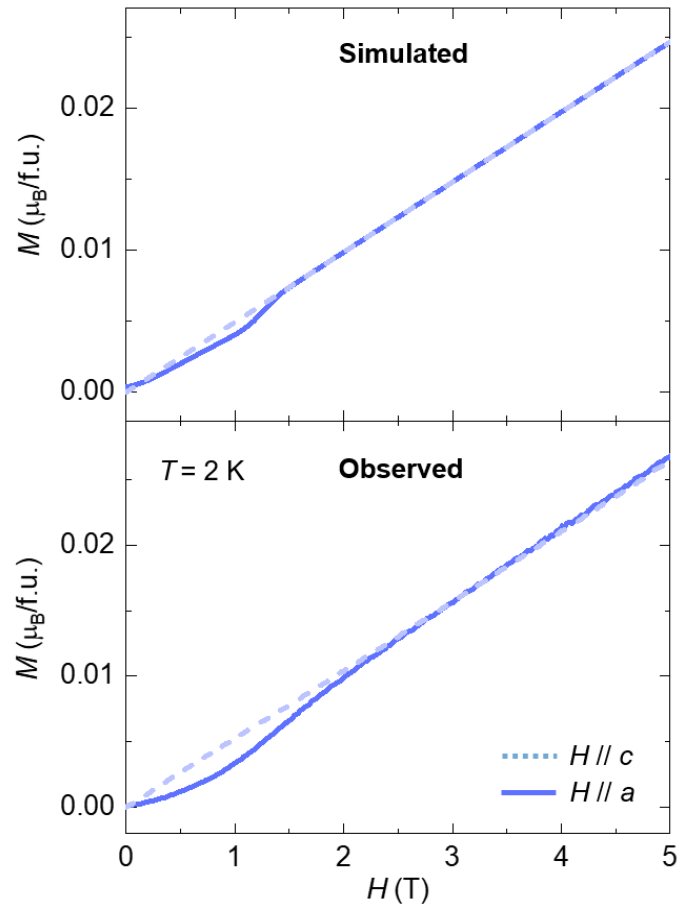

**Fig. S4 Isothermal magnetization along *a* and *c* axes at *T* = 2 K.** Comparison of measured (bottom) and simulated (top) $M_a$ (solid lines) and $M_c$ (dashed lines). The simulations use the same parameters as in Fig. 3 with 20% minor twin contribution, accounting for spin chains rotated toward the *a* axis.

**References**

1. Heinrich, M. *et al.* Structural and magnetic properties of $CuSb_2O_6$ probed by ESR. *Physical Review B* **67**, 224418 (2003).

2. Herak, M., Žilić, D., Matković Čalogović, D. & Berger, H. Torque magnetometry study of magnetically ordered state and spin reorientation in the quasi-one-dimensional *S* = 1/2 Heisenberg antiferromagnet $CuSb_2O_6$. *Physical Review B* **91**, 174436 (2015).
3. Nakua, A., Yun, H., Reimers, J., Greedan, J. & Stager, C. Crystal structure, short range and long range magnetic ordering in $CuSb_2O_6$. *Journal of solid state chemistry* **91**, 105-112 (1991).
4. Giere, E.-O., Brahimi, A., Deiseroth, H. & Reinen, D. The Geometry and Electronic Structure of the $Cu^{2+}$ Polyhedra in Trirutile-Type Compounds $Zn(Mg)_{1-x}Cu_xSb_2O_6$ and the Dimorphism of $CuSb_2O_6$: A Solid State and EPR Study. *Journal of Solid State Chemistry* **131**, 263-274 (1997).
5. Maimone, D. T., Christian, A., Neumeier, J. & Granado, E. Coupling of phonons with orbital dynamics and magnetism in $CuSb_2O_6$. *Physical Review B* **97**, 174415 (2018).
6. Bonner, J. C. & Fisher, M. E. Linear magnetic chains with anisotropic coupling. *Physical Review* **135**, A640 (1964).
7. Goodenough, J. B. Theory of the role of covalence in the perovskite-type manganites [La, M(II)]$MnO_3$. *Physical Review* **100**, 564 (1955).
8. Kanamori, J. Superexchange interaction and symmetry properties of electron orbitals. *Journal of Physics and Chemistry of Solids* **10**, 87-98 (1959).